\documentclass[11pt]{article} 

\usepackage[utf8]{inputenc} 

\usepackage{geometry} 
\usepackage{graphicx} 

\usepackage{booktabs} 
\usepackage{array} 
\usepackage{paralist} 
\usepackage{verbatim} 
\usepackage{subfig} 
\usepackage[version=3]{mhchem}
\usepackage{here}

\usepackage{fancyhdr} 
\usepackage{sectsty}
\allsectionsfont{\sffamily\mdseries\upshape} 

\usepackage[nottoc,notlof,notlot]{tocbibind} 
\usepackage[titles,subfigure]{tocloft} 

\title{Disappearance of spin glass behavior in \ce{ThCr2Si2}-type intermetallic \ce{PrAu2Si2}}
\author{D. X. Li, Y. Shimizu, A. Nakamura, Y. J. Sato, A. Maurya, Y. Homma,\\ F. Honda, and D. Aoki\\
\\Institute for Materials Research, Tohoku University, Oarai, Ibaraki 311-1313, Japan}

\begin{document}
\maketitle

\begin{abstract}
It is unexpected that a spin-glass transition, which generally occurs only in the system with some form of disorder, was observed in the \ce{ThCr2Si2}-type compound \ce{PrAu2Si2} at a temperature of $\sim$3 K. This puzzling phenomenon was later explained based on a novel dynamic frustration model that does not involve static disorder. We present the results of re-verification of the reported spin-glass behaviors by measuring the physical properties of three polycrystalline \ce{PrAu2Si2} samples annealed under different conditions. Indeed, in the sample annealed at 827 $^\circ$C for one week, a spin-glass transition does occur at a temperature of $T_\mathrm{f}$ $\sim$ 2.8 K as that reported previously in the literature. However, it is newly found that the spin-glass effect is actually more pronounced in the as-cast sample, and almost completely disappears in the well-annealed (at 850 $^\circ$C for 4 weeks) sample. The apparent sample dependence of the magnetic characteristics of \ce{PrAu2Si2} is discussed by comparing it with similar phenomena observed in the isomorphic compounds \ce{URh2Ge2} and \ce{CeAu2Si2}. Our experimental results strongly suggest that the spin-glass behavior observed in the as cast and insufficient annealed samples is most likely due to the presence of small amount of crystalline impurities and/or partial site disorder on the Au and Si sublattices, and thus is not the inherent characteristic of ideal \ce{ThCr2Si2}-type \ce{PrAu2Si2}. The perfectly ordered \ce{PrAu2Si2} should be regarded as a paramagnetic system with obvious crystal-field effects. 

\end{abstract}
\section{Introduction}
Since the discovery of the spin-glass (SG) effect in dilute magnetic systems, the SG phenomenon has been studied experimentally and theoretically for more than 40 years.  So far, a large number of SG materials with different structural characteristics have been discovered. As typical examples, following two series of SG materials are well known.  One is the series of the SG materials in dilute magnetic system (such as Au-Fe or Cu-Mn alloys) [1, 2], where the distribution of dilute magnetic atoms in nonmagnetic metals is geometrically disordered. The other is those in nonmagnetic atom disordered system (such as $R_{2}MX_{3}$, $R$=U or rare earth element, $M$=transition metal, $X$=Si, Ge, Sn, etc.) [3, 4], where the magnetic atoms form an ordered sublattice structure, but the distribution of the nonmagnetic atoms (at least part of them) is chemically disordered at the sublattice sites. In both cases, disordered distribution of atoms can produce random bonds between the magnetic atoms, resulting in the randomly distributed frustrated magnetic moments. Nowadays, in this research field, the relationship between the form of disordered structure and the driving mechanism of SG behavior is still receiving extensive attention. It is generally accepted that ``randomness" and ``frustration" are necessary conditions for generating the SG state, and the SG state cannot be formed in a crystallographically ordered system [5].

The ternary intermetallic compound \ce{PrAu2Si2} appears to be a latent exception. The stoichiometric \ce{PrAu2Si2} crystallizes in a perfectly ordered \ce{ThCr2Si2}-type tetragonal crystal structure, in which Pr, Au, and Si atoms are periodically arranged. The calculation results based on the intensity of x-ray pattern indicate that any random distribution between Au and Si atoms can be excluded [6, 7]. It is therefore surprising that this system was found to show the SG freezing at $T_\mathrm{f}$ $\sim$ 3 K [8, 9]. The typical SG behaviors reported in the literature include the frequency-dependent peak in ac susceptibility around $T_\mathrm{f}$  and the inconsistent temperature dependence of field-cooled (FC) and zero-field-cooled (ZFC) dc magnetization below $T_\mathrm{f}$, etc. In addition, no magnetic Bragg peaks were observed from the neutron powder diffraction experiments down to 1.5 K [9], indicating the absence of long-range magnetic order. The polycrystalline \ce{PrAu2Si2} sample used in these measurements was annealed at 827 $^\circ$C for one week. It is believed that a small amount of Au and Si atoms in this sample may occupy other sites of the \ce {ThCr2Si2} type structure and cause the observed SG effect [8].  However, $\mu$SR spectroscopy down to 40 mK shows no evidence that the SG state is formed in \ce{PrAu2Si2} [10]. Moreover, the subsequent Mössbauer studies down to 1.7 K along with neutron and x-ray diffraction measurements [11] also exclude both SG ordering and collinear antiferromagnetic order of the Pr moments. Meanwhile, the site disorder of Au/Si in the \ce{PrAu2Si2} sample used in the work reported in Ref. 9 was determined to be less than 1\%, which is much smaller than the initially estimated value [8]. Then, the important question is why the bulk measurements on the \ce{PrAu2Si2} sample with so few disordered atoms exhibit the SG features, and how to understand the inconsistent experimental results of the bulk measurements and the $\mu$SR spectroscopy?  If the SG behavior is proved to be an inherent characteristic of perfectly ordered \ce{PrAu2Si2} and independent of structural disorder, the mechanism of the SG phenomenon should be reconsidered from a new physical point of view. Interestingly, Ryan {\it et al}. \cite{Ref-11} suggested that the susceptibility peak observed in the \ce{PrAu2Si2} sample is likely a dynamic effect. Later on, Goremychkin {\it et al}. proposed a novel dynamic frustration model (that do not involve static disorder) \cite{Ref-12} to explain the SG phenomenon in \ce{PrAu2Si2}. Based on this model, the SG behavior in \ce{PrAu2Si2} can be explained by the dynamic fluctuations of the crystal-field (CF) levels, because this system has a singlet ground state and the interionic exchange coupling is very close to the critical value to induce magnetic order.

The striking results between the bulk measurements and the  $\mu$SR/Mössbauer spectroscopy mentioned above motivated us to carefully investigate the \ce{PrAu2Si2} polycrystals annealed under different conditions. The main objectives of this study are (1) to verify the reproducibility of the reported SG behavior, (2) to explain the origin of SG phenomena observed by bulk measurements, and (3) to reveal the actual nature of the ground state of \ce{PrAu2Si2}. In this article, we present the experimental results of ac and dc susceptibility, specific heat and electrical resistivity measurements of three polycrystalline \ce{PrAu2Si2} samples, and discuss the influence of heat treatment on their magnetic properties. The obtained results are compared with those reported in the literature for the isomorphic compounds \ce{URh2Ge2} and \ce{CeAu2Si2}.

\section{Experiment details}

The polycrystalline \ce{PrAu2Si2} sample was prepared by arc melting high-purity raw metals [Pr: 99.9\% (3N); Au: 4N; Si: 6N] in a stoichiometric ratio under a purified argon atmosphere. The melted button was flipped and remelted four times to ensure homogeneity. Several pieces cut from the arc melted ingot were separately wrapped into different tantalum foils and annealed in evacuated quartz tubes under different conditions. The crystal structure and quality of the samples were checked by powder x-ray diffraction (XRD) at room temperature. In this study, following three polycrystalline \ce{PrAu2Si2} samples were selected for use: (1) the as-cast sample labeled S1, (2) the sample annealed at 827 $^\circ$C for one week (the same heat treatment conditions as reported in Refs. [8] and [9]) labeled S2, and (3) the sample annealed at 850 $^\circ$C for 4 weeks labeled S3. The dc magnetization and ac susceptibility were measured at temperatures between 2 and 300 K and  between 2 and 10 K by using a 5 T- and a 7 T-SQUID (superconducting quantum interference device) magnetometer (Quantum Design, Inc.), respectively. Specific-heat measurements were performed by a thermal-relaxation technique in the temperature range of 0.4--20 K employing a physical properties measurement system (PPMS, Quantum Design, Inc.). Electrical resistivity was measured at temperatures between 2 and 300 K using a standard four-terminal ac method.

\section{Results and discussion}

Figure 1 shows the powder XRD patterns of samples S1, S2, and S3 measured at room temperature. For all three samples, the observed main peaks can be indexed based on the symmorphic tetragonal \ce{ThCr2Si2}-type structure (space group $I4/mmm$, \#139, $D ^{17}_{4h}$). In this structure, Pr, Au, and Si atoms distribute in order at the Wyckoff positions 2$a$ (0, 0, 0), 4$d$ (0, 1/2, 1/4), and 4$e$ (0, 0, $z$), respectively \cite{Ref-13}. Using the integrated x-ray powder diffraction software PDXL2 (Rigaku Corporation), the  lattice parameters are determined to be $a$=4.315 (3), 4.3038(13), and 4.3057(16) {\AA}, and $c$=10.245(9), 10.2186(4), and 10.223(5) {\AA} for S1, S2, and S3, respectively, which are in good agreement with the previously reported values [8, 9]. Figure 1 also shows the Bragg positions (see the vertical bar at the bottom of this figure) determined by using lattice constants of S3 and above-mentioned 
\begin{figure}[t]
\begin{center}
\includegraphics[width=8.2cm]{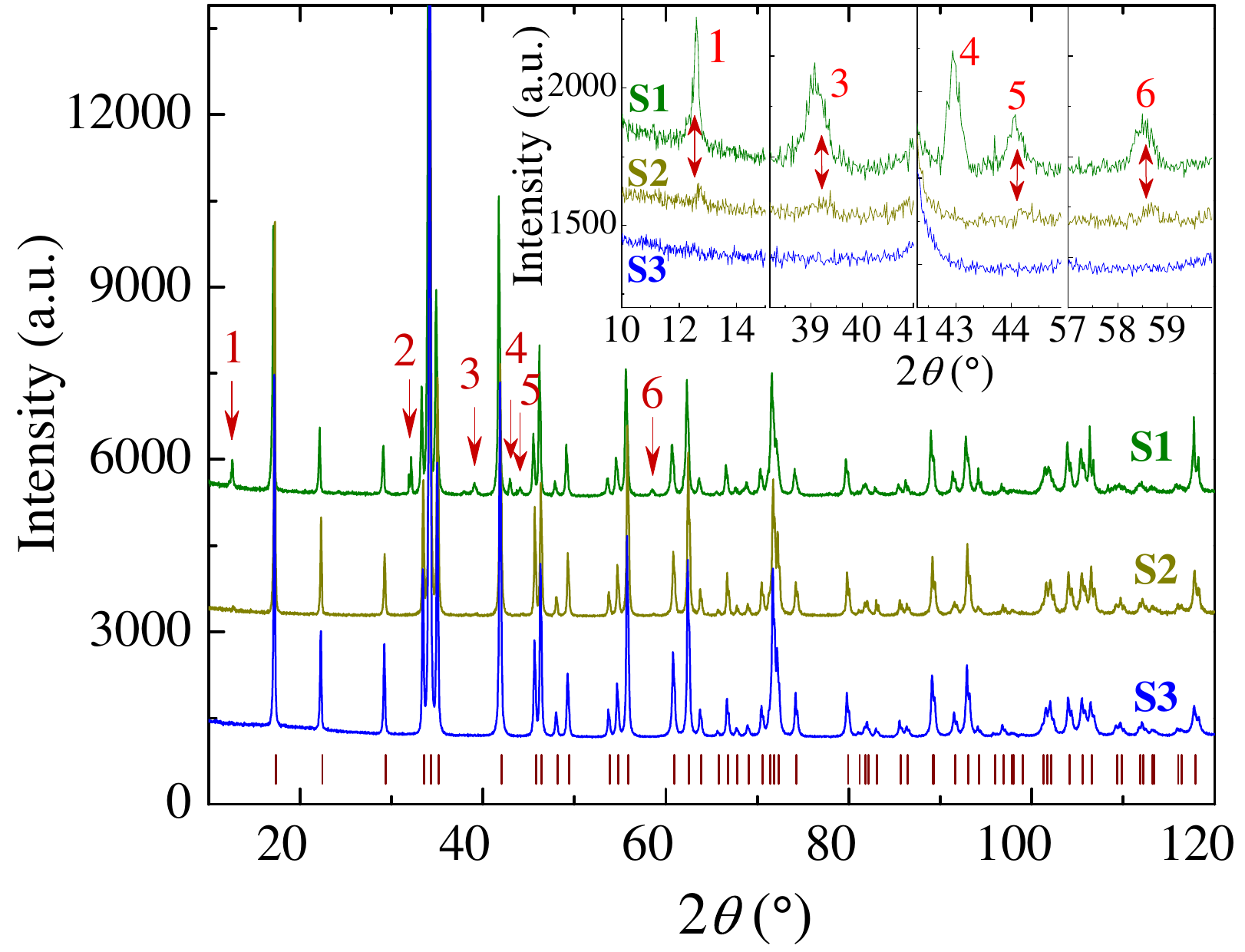}
\end{center}
\caption{
(Color online) Experimental results of powder x-ray diffraction of \ce{PrAu2Si2} samples S1, S2 and S3 at room temperature. The numbered arrows mark the impurity peaks. The vertical bars at the bottom of this figure illustrate the Bragg positions. The inset shows the impurity peaks in the expanded scales.}

\end{figure}
atom coordinates with the refined free parameter $z$=0.368 \cite{Ref-8}. It is clear from Fig. 1 that several small unindexed peaks (marked with numbered arrows) that are not generated from the \ce{ThCr2Si2}-type crystal phase appear in the pattern of S1, indicating the presence of impurities in the as-cast sample. On the other hand, it is clear from the enlarged views shown in the inset of Fig. 1 that these impurity peaks are significantly weakened in the pattern of S2 and almost completely disappeared in the pattern of S3. Using the above-mentioned lattice parameters of sample S1, we have also calculated the Bragg positions of \ce{CeBe2Ge2}-type (space group $P4/nmm$) \ce{PrAu2Si2}, which is considered to possibly exist in S1 in small amounts (for the sake of clarity, the calculated results are not shown). However, at least the two largest impurity peaks (marked by arrows 1 and 2 in Fig. 1) cannot be explained by the  \ce{CeBe2Ge2}-type structure. Although we are currently unable to quantitatively determine the impurity composition and impurity level in S1 at this stage,  the XRD  results show in Fig. 1 clearly indicate that the impurities observed in S1 are partly removed in S2 and almost completely taken away in S3. That is, S1 is a relatively poor quality sample that contains more impurities. S2 is better than S1, and S3 is a relatively high-quality sample, which is better than both S1 and S2. With this information in mind, we will reveal the different magnetic properties of the three samples below.


\begin{figure}[t]
\begin{center}
\includegraphics[width=8.6cm]{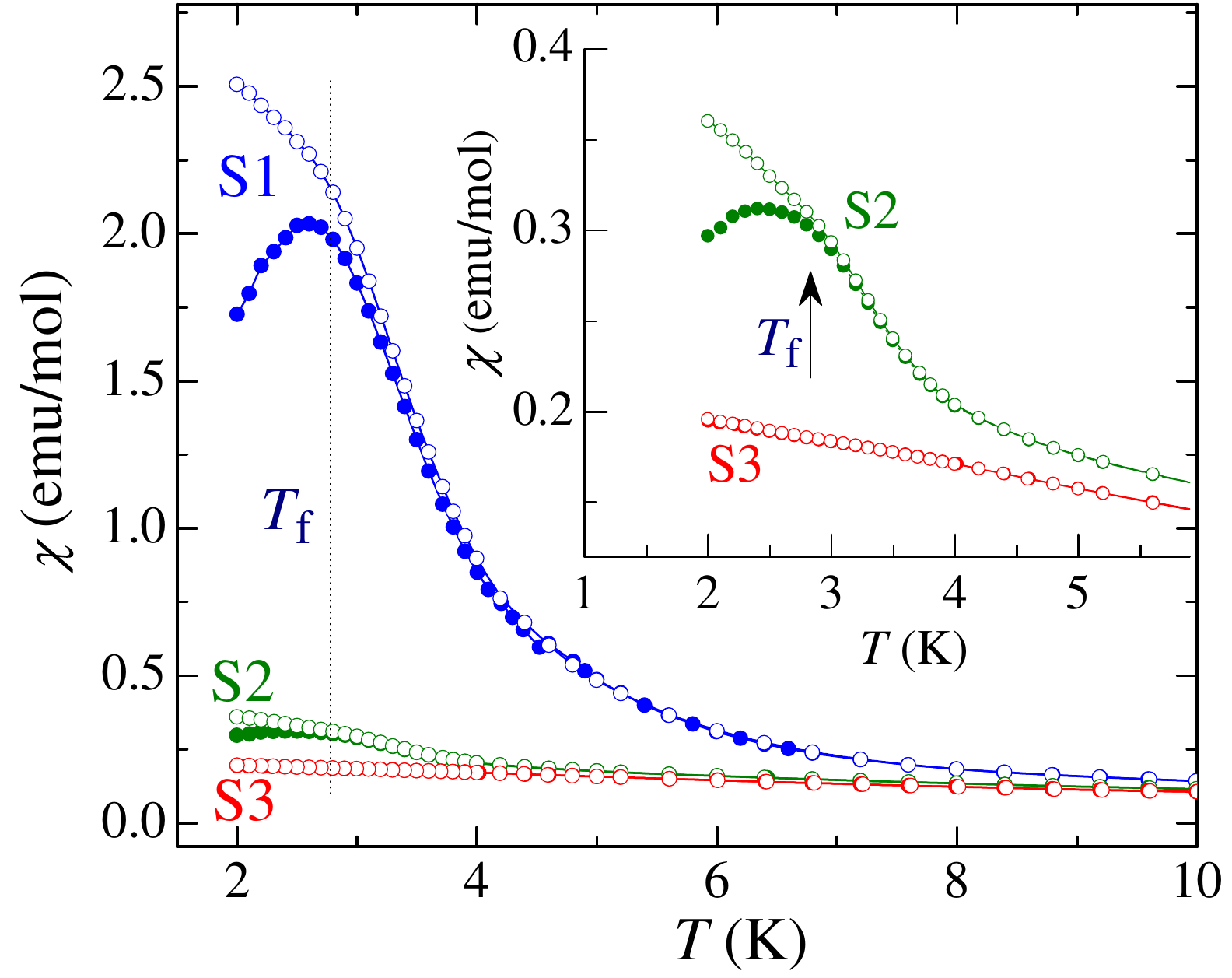}
\end{center}
\caption{
(Color online) Temperature dependences of the dc susceptibility, $\chi(T)$, of the \ce{PrAu2Si2} samples S1, S2 and S3 measured with FC ($\circ$) and ZFC ($\bullet$) processes in a field of 10 Oe. Inset shows the data of samples S2 and S3 in an expanded scale.}

\end{figure}

The temperature ($T$) and magnetic field ($H$) dependences of magnetization ($M$) of samples S1, S2, and S3 are measured with the FC and the ZFC processes. Figure 2 displays the results of susceptibility ($\chi=M/H$) below 10 K at $H$=10 Oe. The data of samples S2 and S3 are also shown in the inset of this figure in an expanded scale. Clearly, samples S1, S2 and S3 exhibit significantly different $\chi(T)$ behavior. For sample S2, the $\chi_\mathrm{ZFC}(T)$ curve shows a cusp just below $T_\mathrm{f}$ $\sim$ 2.8 K, and evident difference between the FC and ZFC curves appears below the cusp temperature [see the inset of Fig. 2]. This behavior is analogous to that reported in the literature [8, 9], and represents the typical feature of SG materials [5, 14-16]. In fact, it is clear from Fig. 2 that the SG behavior is more pronounced in the as-cast sample S1 manifesting as a much larger peak in the $\chi_\mathrm{ZFC}(T)$ curve. This result suggests that more Pr atoms in the as-cast sample participate in the SG state. A more important finding in this study is that the $\chi_\mathrm{ZFC}(T)$ curve of sample S3 does not show cusp around $T_\mathrm{f}$ $\sim$ 2.8 K, both  $\chi_\mathrm{FC}(T)$ and  $\chi_\mathrm{ZFC}(T)$ of S3 increase smoothly with decreasing $T$ down to 2 K, and there is almost no difference between them [see the inset in Fig. 2]. This result means that the SG effect observed in S1 and S2 disappears in S3.

\begin{figure}[t]
\begin{center}
\includegraphics[width=8.7cm]{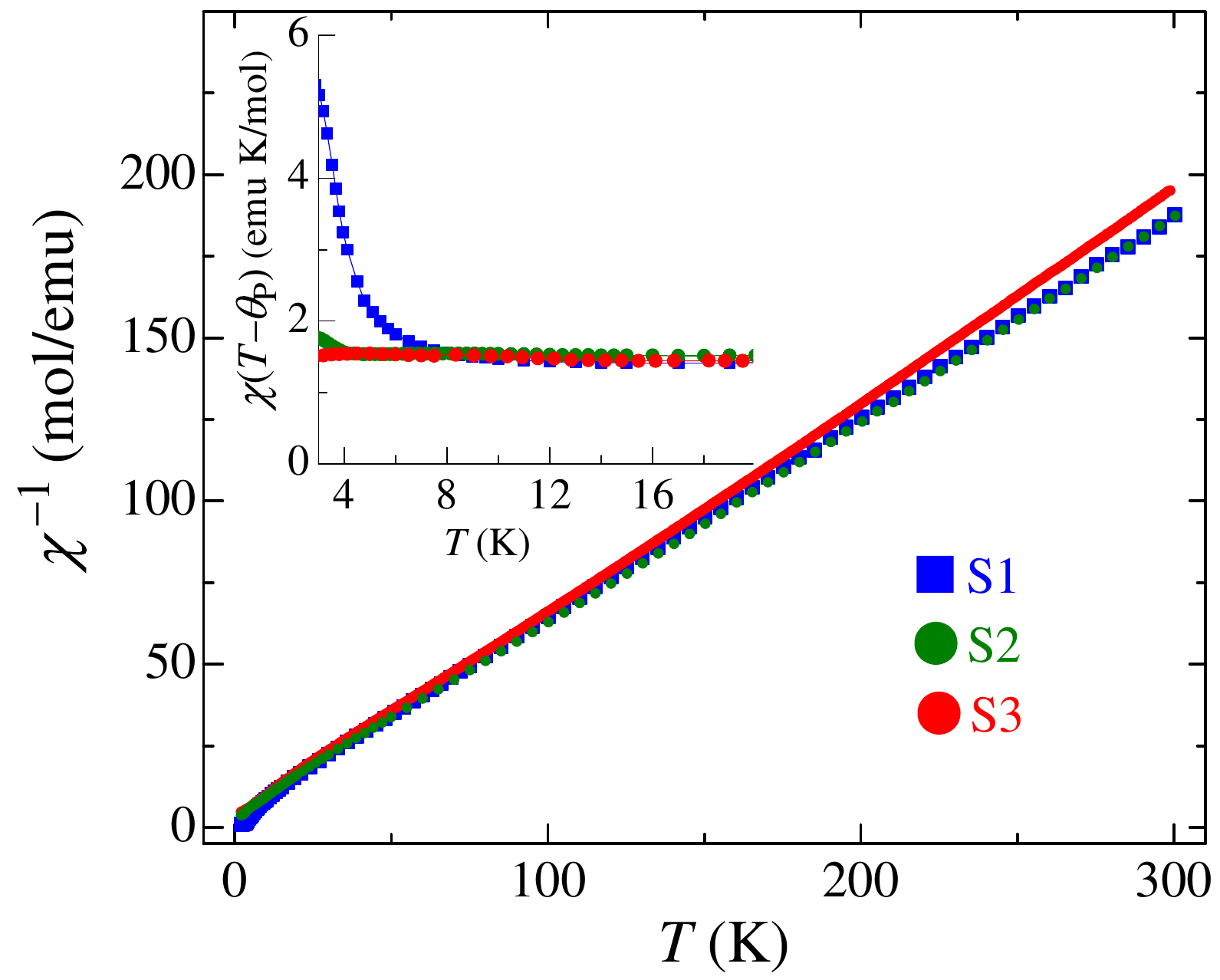}
\end{center}
\caption{
(Color online) Temperature dependences of the inverse susceptibility, $\chi(T)^{-1}$, of the \ce{PrAu2Si2} samples S1, S2 and S3 measured in a field of 100 Oe with ZFC processes. Inset displays the plot of  $\chi (T-\theta_\mathrm{P})$ vs $T$ between 3 and 20 K.}

\end{figure}

Figure 3 shown the temperature dependence of  inverse susceptibility, $\chi(T)^{-1}$, of the \ce{PrAu2Si2} samples S1, S2 and S3 measured in a field of 100 Oe with ZFC processes. At high temperatures, $\chi(T)$ of the three samples reveals similar Curie-Weiss behavior. Fitting the data above 100 K to Curie-Weiss law yields the effective magnetic moment $\mu_\mathrm{eff}$ = 3.61, 3.58, and 3.50 $\mu_\mathrm{B}$/Pr, and the paramagnetic Curie temperature $\theta_\mathrm{P}$ = 4.5, -0.4, and -0.7 K for S1, S2, and S3, respectively. The determined $\mu_\mathrm{eff}$ values are in good agreement with that expected for a free Pr$^{3+}$ ion with an $4f ^{2}$ electronic configuration, while the small  $\theta_\mathrm{P}$ values suggest the weak magnetic exchange interactions. As the temperature decreases, $\chi(T)$ starts to deviate from the Curie-Weiss law at about 100 K, which may be due to the CF effect. When the temperature is lowered further, $\chi(T)$ again approximately follows the Curie-Weiss law between about 14 and 22 K, which yields $\theta_\mathrm{P}$ =-2.4, -4.4, and -4.7 K, and  $\mu_\mathrm{eff}$ =3.36, 3.48  and 3.40 $\mu_\mathrm{B}$/Pr for S1, S2, and S3, respectively. The inset in Fig. 3 shows the behavior of  ``Curie constant'' [=$\chi (T-\theta_\mathrm{P}$)] between 3 and 20 K. For sample S3 with no SG transition, $\chi (T-\theta_\mathrm{P})$ remains almost constant. In contrast, for samples S1 and S2 with clear SG transitions, the $\chi (T-\theta_\mathrm{P})$ value begins to increase obviously with further decreasing $T$ to about 8 K and 4 K, respectively. Which suggests that  $\chi(T)$ behavior again deviates from the Curie-Weiss law due to the SG correlations starting to develop from the paramagnetic background. 
Especially for sample 1, the increase in $\chi (T-\theta_\mathrm{P})$ value is more significant. Note that similar experimental phenomenon has been reported for a nonmagnetic atom disorder SG system \ce{Ce2AgIn3} [17].

\begin{figure}[t]
\begin{center}
\includegraphics[width=8.7cm]{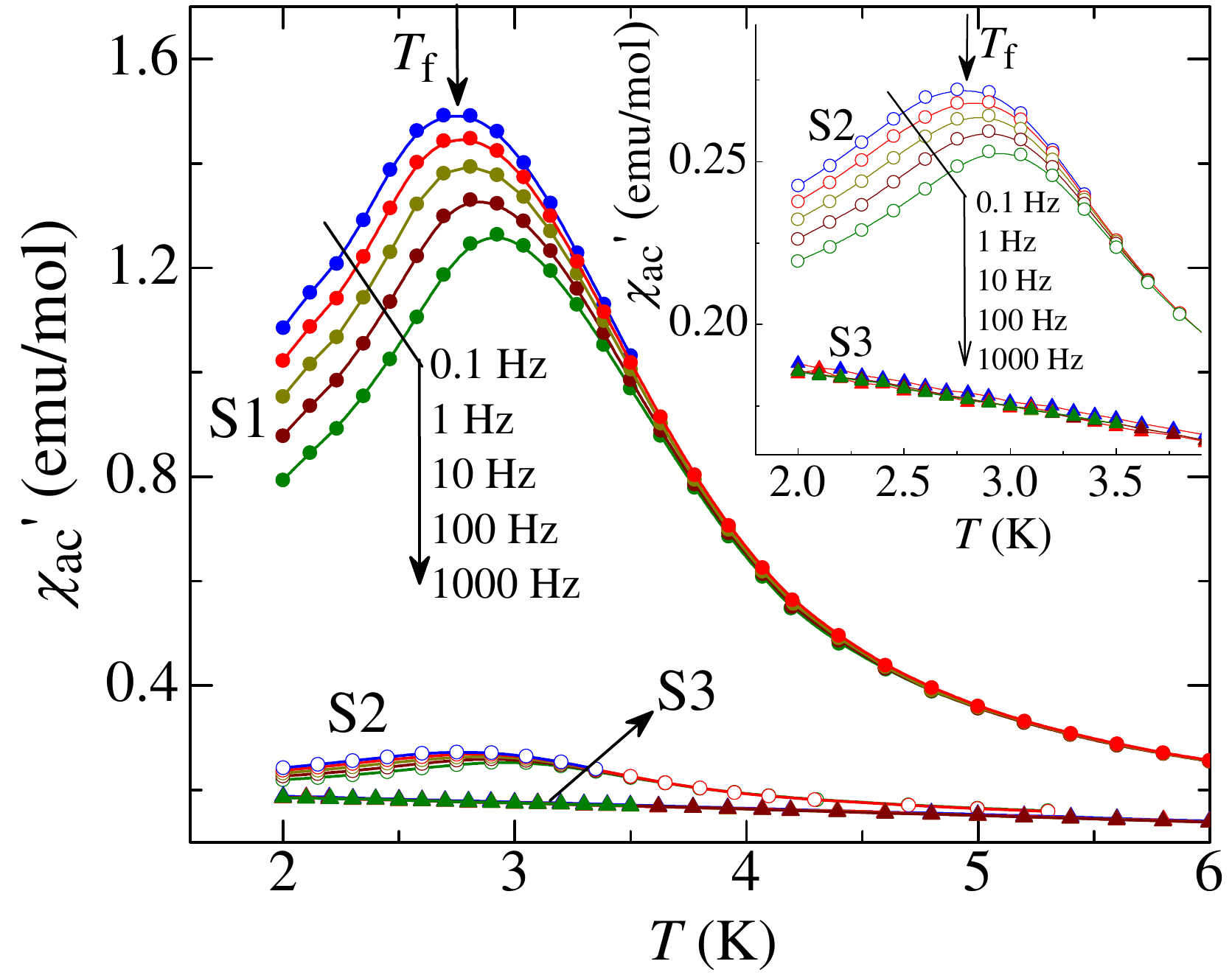}
\end{center}
\caption{
(Color online) Temperature dependences of the real  component of the ac susceptibility [$\chi '_\mathrm{ac}$($T,\omega$) ] of \ce{PrAu2Si2} samples S1, S2 ans S3 measured at 0.1, 1, 10, 100 and 1000 Hz in an ac field of 1 Oe. The inset shows the $\chi '_\mathrm{ac}$($T,\omega$) data of samples S2 and S3 below 4 K in an expanded scale.}

\end{figure}

To further confirm the SG effect,  ac susceptibility measurements were performed at the frequency range of 0.1 Hz $\leq\omega/2\pi\leq$1000 Hz. The in-phase components $\chi '_\mathrm{ac}$($T,\omega$) of the ac susceptibility of the the three samples between 2 and 6 K are presented in Fig. 4, while the curves of S2 and S3 are enlarged and shown in the inset of this figure.  For S2, the $\chi '_\mathrm{ac}$($T,\omega$)  curve shows a noticeable maximum near $T_\mathrm{f}$ $\sim$ 2.8 K, and its amplitude and position depend on the frequency $\omega$ of the applied ac magnetic field. This is also a typical feature of SG materials [5, 14, 18-20] and is consistent with the results reported in the literature [8, 9]. In fact, it is clear from Fig. 4 that a similar frequency-dependent ac-susceptibility peak is more pronounced in S1 indicating much stronger SG effect in the as-cast sample. Defining the peak temperature in $\chi '_\mathrm{ac}$($T$) as the SG transition point $T_\mathrm{f}$, it is observed that there is almost no difference in $T_\mathrm{f}$  value between S1 and S2. At $\omega/2\pi$ = 0.1 Hz, $T_\mathrm{f}$  is determined to be $\sim$2.79 K, which shifts to $\sim$2.95 K at $\omega/2\pi$ = 1000 Hz. The corresponding rate of the initial frequency shift, defined as  
$\delta T_\mathrm{f}$ = $\bigtriangleup T_\mathrm{f}/(T_\mathrm{f}\bigtriangleup\mathrm{log}\omega)$,
is $\delta T_\mathrm{f}\sim0.014$, which is comparable to the typical value (from a few thousandths to a few percent) of most spin glasses [4, 5, 21-23]. These  results give further evidence for the SG state in S1 and S2.  On the other hand, it is particularly important that $\chi '_\mathrm{ac}$($T,\omega$) of the sample S3 increases smoothly as $T$ decreases down to 2 K without any anomaly and frequency dependence around $T_\mathrm{f}\sim$ 2.8 K. This finding clearly indicates that no SG state is formed in S3.

The apparent sample dependence of low-temperature magnetic properties of \ce{PrAu2Si2} is reminiscent of the similar phenomenon reported for two isomorphic (\ce{ThCr2Si2}-type) compounds \ce{URh2Ge2} and \ce{CeAu2Si2}. Single crystal \ce{URh2Ge2} in as-grown form is reported to show evidence of a SG order below $T_\mathrm{f}$ = 11 K, with irreversible magnetism and the frequency-dependent ac susceptibility [23]. However, the SG effect is attenuated in the heat-treated samples, in particular, annealing at temperatures of the order of 1000 $^\circ$C is sufficient to transform the \ce{URh2Ge2} sample from a SG system into a long-range ordered antiferromagnet [24]. Formation of SG state in the as-grown single crystal \ce{URh2Ge2} is thus considered to originate from some site disorder on the Rh and Ge sublattices due to a possible amalgamation of the \ce{ThCr2Si2} and the \ce{CeBe2Ge2} structures. 

\begin{figure}[htbp]
\begin{center}
\includegraphics[width=8.8cm]{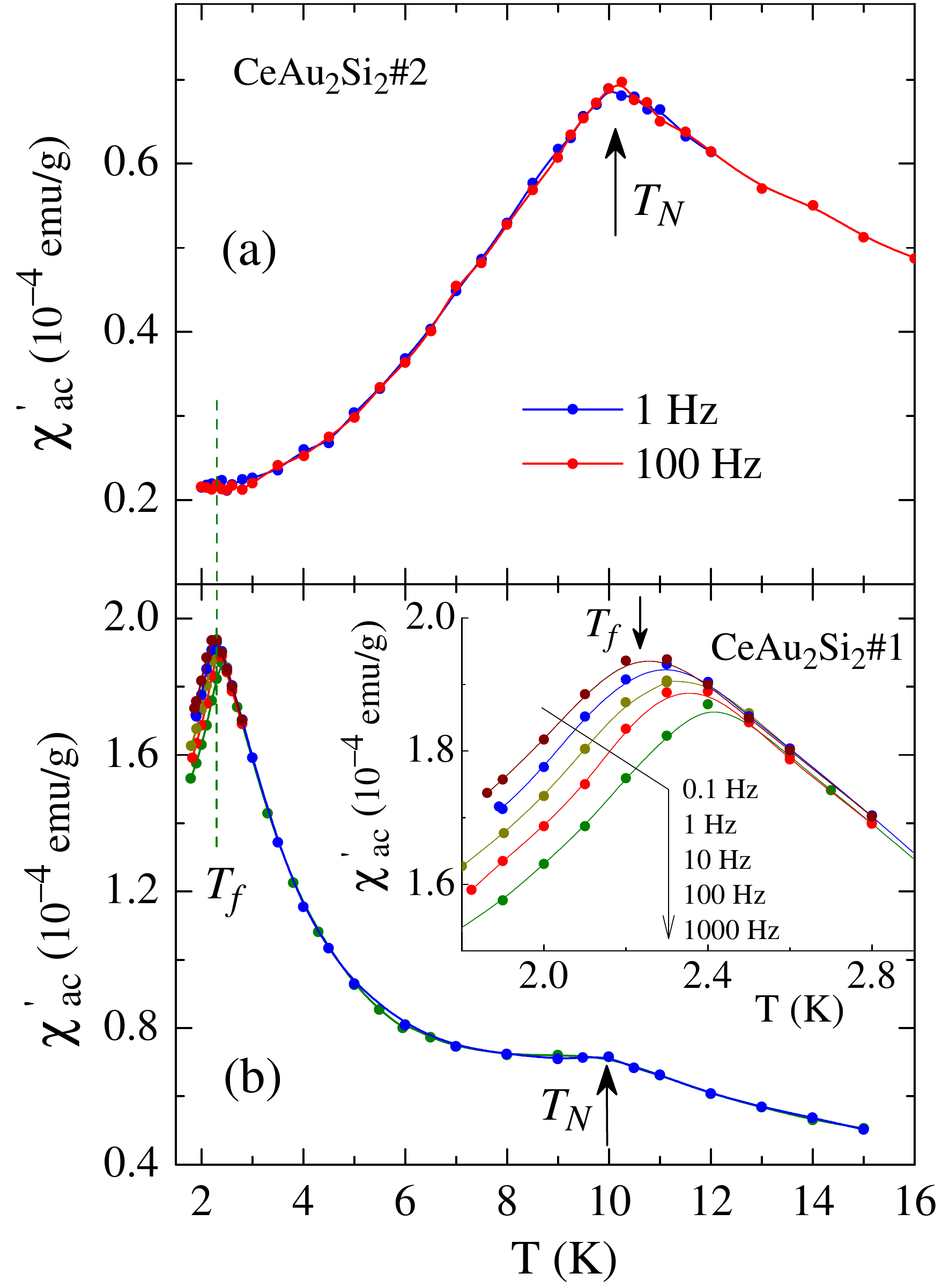}
\end{center}
\caption{
(Color online)  Temperature dependences of the real  component of the ac susceptibility [$\chi '_\mathrm{ac}$($T,\omega$) ] of \ce{CeAu2Si2}\#2 (a) and  \ce{CeAu2Si2}\#1 (b)  at various frequencies in an ac field of 1 Oe. The inset in  (b) displays the expanded plots of the $\chi '_\mathrm{ac}$($T,\omega$) data of \ce{CeAu2Si2}\#1 around $T_\mathrm{f}$ $\sim$2.2 K. The data of \ce{CeAu2Si2}\#1 has been published in Ref. [16], and show here again for comparison with \ce{CeAu2Si2}\#2.}

\end{figure}

In addition, very similar phenomenon was also observed in \ce{CeAu2Si2}. In an earlier paper [25], we reported the classic SG behavior (irreversible magnetism and frequency-dependent ac susceptibility, etc.) in a polycrystalline \ce{CeAu2Si2} annealed at 800 $^\circ$C for 2 weeks (marked as \ce{CeAu2Si2}\#1 in this paper), which shows a SG transition at $T_\mathrm{f}$ $\sim$ 2.2 K much lower than its N\'eel point $T_\mathrm{N}$ = 10 K. However, our recent experimental results reveal that the above-mentioned SG behaviors in \ce{CeAu2Si2}\#1 completely disappear in  a polycrystalline sample annealed at 850 $^\circ$C for 39 days (marked as \ce{CeAu2Si2}\#2 in this paper). Note that both \ce{CeAu2Si2}\#1 and \ce{CeAu2Si2}\#2 samples were cut from the same arc melted ingot. Figure 5(a) displays the results of real component of the ac susceptibility measurements of \ce{CeAu2Si2}\#2. The early published data for \ce{CeAu2Si2}\#1  [25] are also shown here for comparison [Fig. 5(b)]. Clearly, the frequency-dependent large peak observed in $\chi '_\mathrm{ac}$($T,\omega$) curve of \ce{CeAu2Si2}\#1 near $T_\mathrm{f}$ $\sim$ 2.2 K, which represents the SG transition, disappears completely in the $\chi '_\mathrm{ac}$($T,\omega$) curve of \ce{CeAu2Si2}\#2. This result definitely signifies that no SG state is formed in \ce{CeAu2Si2}\#2. The formation of SG state in \ce{CeAu2Si2}\#1 may also be attributed to the crystal inhomogeneity due to insufficient annealing. Compared to \ce{CeAu2Si2}\#1, \ce{CeAu2Si2}\#2 seems to be a well-annealed sample with much better crystal uniformity.  As a matter of fact, recent studies on single crystalline \ce{CeAu2Si2} have further confirmed the absence of SG behavior in high-quality samples [26, 27]. It should be emphasized that at the N\'eel temperature $T_\mathrm{N}$, a frequency-independent ac susceptibility peak can still be observed in the insufficiently annealed sample \ce{CeAu2Si2}\#1, although the amplitude is greatly reduced compared with the well-annealed sample \ce{CeAu2Si2}\#2. In the present case, however, even for the well-annealed \ce{PrAu2Si2} sample S3 showing negative $\theta_\mathrm{P}$, the $\chi '_\mathrm{ac}$($T,\omega$) curve shows no signs of such ambiguous peak, indicating the absence of a similar long-range magnetic order in this system.

In analogy with \ce{URh2Ge2} and \ce{CeAu2Si2}, the differences in the magnetic behavior observed in our three \ce{PrAu2Si2} samples are naturally considered to associate with their microstructure changes caused by the heat treatments. We would like to point out that two possible mechanisms associated with structural disorder may be responsible for the SG behavior in samples S1 and S2. One is the presence of randomly distributed magnetic impurities, which interact weakly with each other, just like that in the conventional dilute magnetic systems. The other is the possible existence of  a part of Au and Si atoms randomly distributed on the sublattice sites, as in the case of nonmagnetic atomic disorder SG systems. Anyway, the as-cast sample S1 seems to remain more magnetic impurities and/or disordered Au and Si atoms, which results in a relatively strong SG effect. In sample S2, magnetic impurities and/or the sources of site disorder may be partially removed but still exist in small amounts (see the inset of Fig. 1) due to the insufficient heat-treatment. Therefore, although the SG effect is significantly suppressed, it can still be detected. However, sample S3 seems to have been well annealed, in which the crystal impurities and/or the source of site disorder may have been almost completely removed, and therefore no SG state is formed. In this sense, annealing at 850 $^\circ$C for 4 weeks should be a relatively appropriate heat-treatment condition to make the microcrystalline structure of our \ce{PrAu2Si2} sample closer to the ideally ordered \ce{ThCr2Si2}-type. 

Correctly assessing the impurity/defect content of samples S1 and S2 may be a topic worth further study. Analyzing the behavior of Curie constant near the spin freezing temperature (such as that shown in the inset in Fig. 3) may be able to provide some useful information for estimating the fraction of impurity/defect in these samples [17, 28, 29]. As seen in the inset of Fig. 3, $\chi (T-\theta_\mathrm{P})$ of sample S1 strongly deviates from the Curie-Weiss behavior below about 8 K as in the case of  \ce{Ce2AgIn3} [17]. Moreover,  it remains unclear whether the magnitude of the loose spins is the same as that in the high-temperature region above 100 K. Therefore, at this stage, it is quite difficult to estimate the fraction of the loose spins that attributed to the SG transition in  our \ce{PrAu2Si2} samples S1 and S2. To clarify this topic, further experimentation and theoretical research are needed. Nevertheless, the above mentioned experimental results clearly show that the SG effect of  \ce{PrAu2Si2} observed in bulk measurements is due to the heterogeneity of the sample. After the quality of the sample is improved by proper heat treatment, the SG effect is weakened and eventually disappears completely in a sufficiently high-quality sample. 

\begin{figure}[htbp]
\begin{center}
\includegraphics[width=8.2cm]{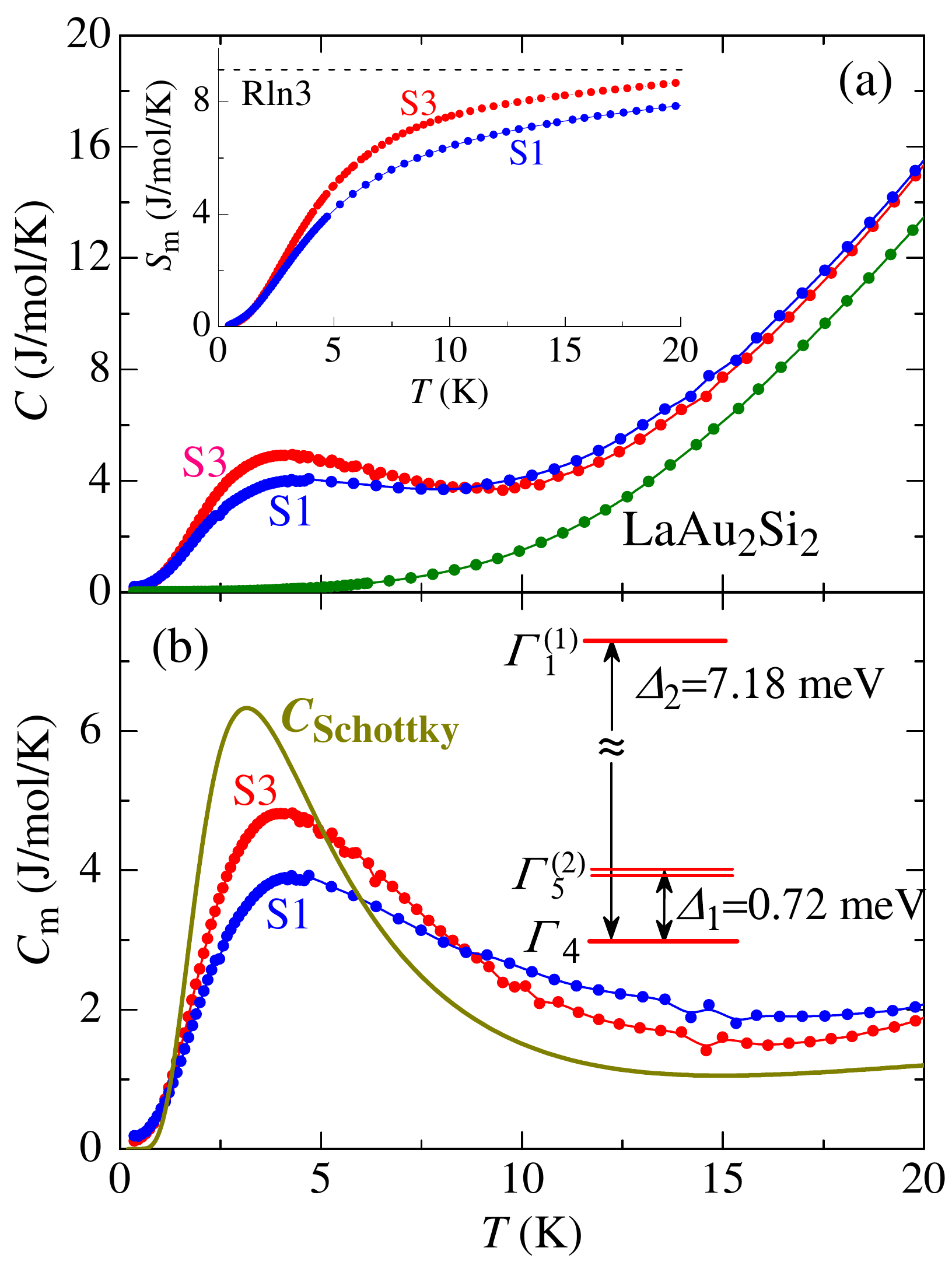} 

\end{center}
\caption{
(Color online) (a) Temperature dependences of specific heat of \ce{PrAu2Si2} samples S1 (with SG effect) ans S3 (without SG effect), and non-{\it f}-electron reference compound LaAu$_{2}$Si$_{2}$. The inset in (a) depicts  the estimated magnetic entropy as a function of temperature for S1 and S3. (b) Temperature dependences of magnetic specific heat of \ce{PrAu2Si2} samples S1 and S3, and the Schottky specific heat calculated by using the CF level parameters shown in the inset of this figure, which is deduced from the inelastic neutron scattering measurements by Goremychkin {\it et al}. [19].}

\end{figure}

To understand the physical properties of  \ce{PrAu2Si2} more comprehensively, we have also measured the temperature dependence of specific heat, $C(T)$, of samples S1 (with SG effect) and S3 (without SG effect) at temperatures between 0.4 and 20 K. The experimental results are presented in Fig. 6(a) together with the data of a non-{\it f}-electron reference compound LaAu$_{2}$Si$_{2}$. We estimate the magnetic specific heat $C_\mathrm{m}(T)$ of \ce{PrAu2Si2} by subtracting the non-magnetic contribution $C(T)_{\mathrm{LaAu_2Si_2}}$ from the total specific heat  $C(T)_{\mathrm{PrAu_2Si_2}}$, which is shown in Fig. 6(b). The magnetic entropy $S_\mathrm{m}(T)$ is then calculated by integrating $C_\mathrm{m}/T$ vs $T$. As displaying in the inset of Fig. 6 (a), magnetic entropy  released up to 20 K (well within the paramagnetic state) is 8.0 and 8.8 J mol$^{-1}$K$^{-1}$, equivalent to 0.87 and 0.96 $R$ln3 for the as cast sample S1 and the well-annealed sample S3, respectively. The entropy value of S1 is slightly smaller than that of S3. This is because the degree of structural order of S1 is lower than that of S3, resulting in a relatively small Schottky specific-heat peak of S1 (see the following). The estimated entropy value of S3 corresponds well to the triple low lying CF levels determined by Goremychkin {\it et al.} based on a inelastic neutron scattering measurement [30]. According to Goremychkin {\it et al.}, the 4$f$ ground state of  \ce{PrAu2Si2} is a $\it{\Gamma}_{\rm 4}$ singlet, and the first excited level is a $\it{\Gamma} ^{(\rm 2)}_{\rm 5}$ doublet at 0.72 meV. The second excited singlet $\it{\Gamma} ^{(\rm 1)}_{\rm 1}$ is located at 7.18 meV, much higher than the first excited energy level [see the inset of Fig. 6(b)]. Another characteristic feature of the specific-heat data is that both S1 and S3 reveal a large and broad peak in $C_\mathrm{m}(T)$ curve around 4 K as illustrated in Fig. 6(b). Since there is no long-range magnetic ordering in \ce{PrAu2Si2} in the experimental temperature range between 0.4 and 300 K, the observed broad peak is most likely due to the Schottky anomaly. We have also calculated the Schottky specific heat, $C(T)_\mathrm{Schottky}$, of \ce{PrAu2Si2} using the above-mentioned CF level parameters. As clearly shown in Fig. 6(b), the $C(T)_\mathrm{Schottky}$ curve exhibits an broad peak near 4 K very similar to that seen in samples S1 and S3, suggesting that the broad specific-heat peak displayed by \ce{PrAu2Si2} near 4 K originates from the Schottky contribution. Comparing with S3, sample S1 shows a relatively small specific heat peak and a relatively large peak width, which may be  caused by the disorder effect.

 At low temperatures, the $C/T$ vs $T^2$ plot shows a linear behavior between 0.6 and 1.5 K [see Fig. 7(a)]. Extrapolating this straight line to $T \rightarrow 0$ yields a value of $\gamma$ = $C/T|_{T \rightarrow 0}$ = 302 mJ (mole Pr)$^{-1}$K$^{-2}$ for S1, which is significantly larger than $C/T|_{T \rightarrow 0}$ = 195 mJ (mole Pr)$^{-1}$K$^{-2}$ for S3. We ascribe such a big difference to the formation of SG state in S1, because the random spin correlations in S1 can increase the contribution of the  $T$-linear term to the specific heat [31]. On the other hand, a nearly linear dependence of $C/T$ vs $T^2$ also appears in the temperature range of 12-20 K (much above the peak temperature) for both S1 and S3, and the extrapolation of this straight line to $T \rightarrow 0$ results in the Sommerfeld coefficients of  $\gamma$ = 202 and 187 mJ (mole Pr)$^{-1}$K$^{-2}$ for S1 and S3, respectively [see Fig. 7(b), only the data above 10 K is shown for the sake of clarity]. As expected, the difference between the $\gamma$ values of S1 and S3 is notably reduced compared to the data obtained from the low temperature $C(T)$ data [see Fig. 7(a)], because the SG correlations in S1 are eliminated at high temperatures. Nevertheless, the $\gamma$ values are still much larger than that expected for a normal intermetallic compound even for the sample S3 with no SG effect.  

\begin{figure}[H]
\begin{center}
\includegraphics[width=8.4cm]{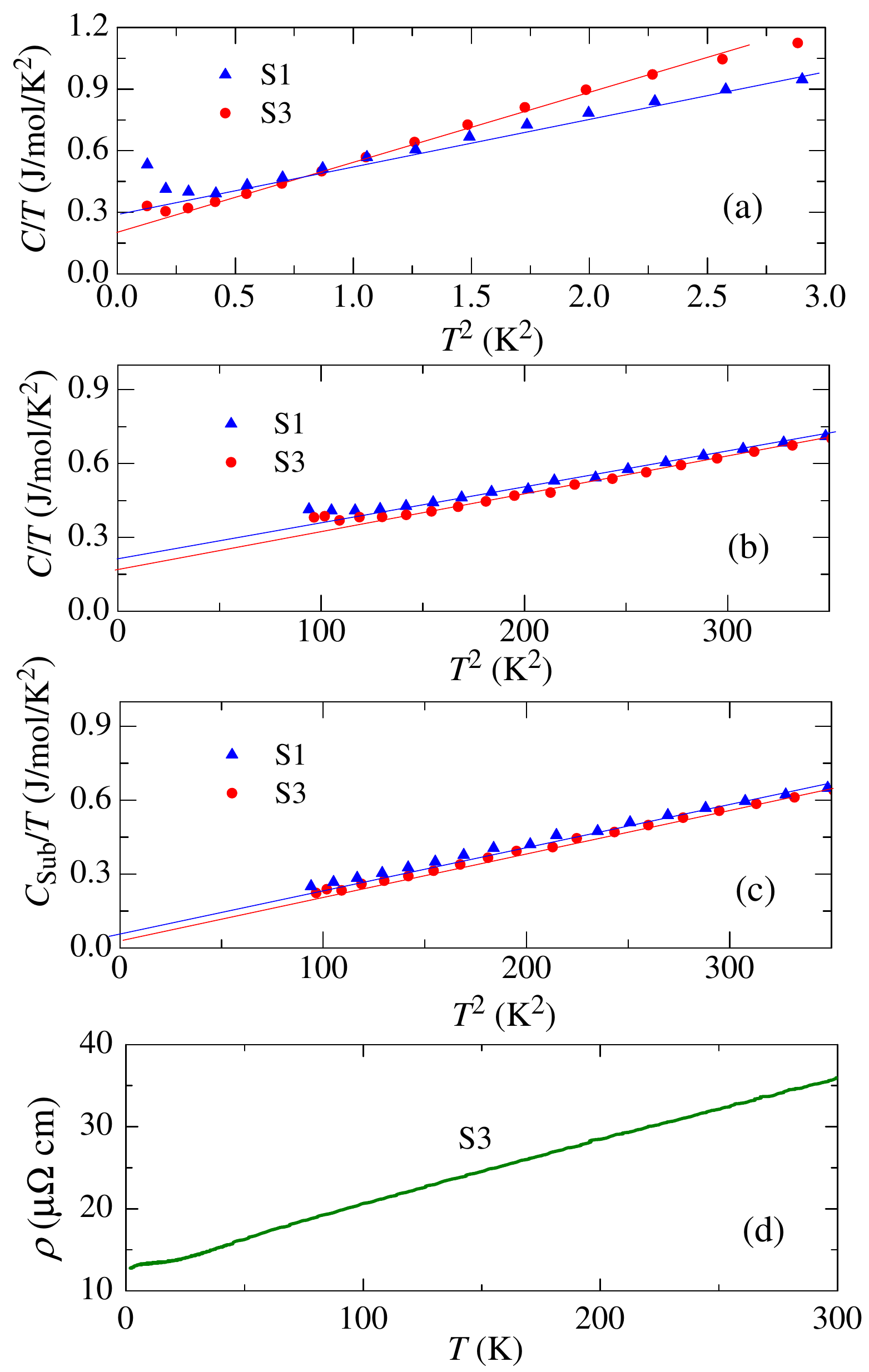}
\end{center}
\caption{
(Color online) (a) $C/T$ vs $T^2$ plots of specific-heat data of the \ce{PrAu2Si2} samples S1 and S3 between 0.4 and 1.7 K. The solid lines represent the linear fits in the range of 0.6-1.5 K. (b) $C/T$ vs $T^2$ plots of specific-heat data of the \ce{PrAu2Si2} samples S1 and S3 between 10 and 20 K. The solid lines represent the linear fits in the range of 12-20 K. (c) $C_\mathrm{Sub}/T$ vs $T^2$ plots [$C_\mathrm{Sub}(T)$ = $C(T)_{\mathrm{PrAu_2Si_2}}$-$C(T)_{\mathrm{Schottky}}$] of specific-heat data of the \ce{PrAu2Si2} samples S1 and S3 between 10 and 20 K. The solid lines represent the linear fits in the range of 12-20 K. (c) Temperature dependence of electrical resistivity of the \ce{PrAu2Si2} sample S3 between 0.4 and 300 K.}

\end{figure}

In general, a large $\gamma$ value is considered to indicate the heavy-fermion behavior. However, given the low-lying CF levels of \ce{PrAu2Si2}, the large Sommerfeld coefficient in this system may be also resulted from the magnetic correlations and/or CF effects [31] as suggested by Krimmel {\it et al}. [8]. As a qualitative analysis, we subtract the Schottky specific heat $C(T)_{\mathrm{Schottky}}$ mentioned above from the total specific heat $C(T)_{\mathrm{PrAu_2Si_2}}$, and using the obtained data $C_\mathrm{Sub}(T)$ [= $C(T)_{\mathrm{PrAu_2Si_2}}$-$C(T)_{\mathrm{Schottky}}$] to re-estimate the $\gamma$ value. Figure 7(c) shows the $C_\mathrm{Sub}/T$ vs $T^2$ plot of S1 and S3, which is also approximately linear in the temperature range of 12-20 K. The  re-estimated $\gamma$ value by extrapolating the straight line to $T \rightarrow 0$ is reduced to 52 mJ (mole Pr)$^{-1}$K$^{-2}$ for S1 and 21 mJ (mole Pr)$^{-1}$K$^{-2}$ for S3. Clearly, the large Sommerfeld coefficient in \ce{PrAu2Si2} is mainly originated from the low-lying CF level effect, and therefore this system cannot be classified as a heavy-fermion material. Similar specific-heat behavior has been observed in the isomorphic compound PrCu$_{2}$Si$_{2}$ [32].  In addition, our electrical resistivity measurements also give no evidence for Kondo effect in \ce{PrAu2Si2}, which shows a metallic $T$-dependence between 2 and 300 K with no long-range magnetic order or superconductivity transition [see Fig.7(d)].

\section{Conclusion}
We have measured the fundamental physical properties of three polycrystalline \ce{PrAu2Si2}  samples annealed under different conditions to verify the presence of SG behavior. As reported in the literature [8], a SG transition did occur in the sample S2 annealed at 827 $^\circ$C for 1 week. In addition, it is newly found that the SG effect in the as-cast sample S1 is actually much stronger than that in S2. Of particular importance is that the SG behavior almost completely disappeared in the sample S3 annealed at 850 $^\circ$C for 4 weeks. The SG behavior observed in the as-cast sample S1 and in the insufficient annealed sample S2 is most likely due to the random distribution of small amount of crystalline impurities and/or partial site disorder on the Au and Si sublattices. By improving the sample preparation method and appropriately selecting the heat-treatment conditions, the crystal structure of \ce{PrAu2Si2} sample can be purified and approached to the ideal ordered \ce{ThCr2Si2}-type. In the purification process of the crystal structure, the random spin correlations in \ce{PrAu2Si2} are gradually suppressed (as in S2) until the SG behavior completely disappears (as in S3). Although we are  currently unable to quantitatively determine the impurity/defects composition and impurity levels in samples S1 and S2,  the XRD patterns clearly show that the three samples have different structural homogeneity. There are more impurities/defects in S1, which are partly removed in S2 and almost completely taken away in S3. According to our experimental results, it can be reasonably considered that the SG state cannot be formed in a \ce{PrAu2Si2} system with a perfectly ordered structure, and thus the SG phenomenon is not an inherent characteristic of the ideal \ce{ThCr2Si2}-type \ce{PrAu2Si2}. On the other hand, the evident anomaly in $C_\mathrm{m}(T)$ around 4 K is most likely due to the Schottky contribution of the low-lying CF levels, which also results in a significant increase in Sommerfeld coefficient. No evidence for long-range ordered magnetic structure, Kondo effect, or superconductivity transition was found in the temperature range between 0.4 and 300 K. It seems much better to classify the perfectly ordered \ce{ThCr2Si2}-type \ce{PrAu2Si2} as a paramagnetic system with obvious CF effects.

\section{Acknowledgments}
A part of the experimental work was performed in the Laboratory of Low Temperature Materials Science, IMR, Tohoku University. The authors gratefully thank
Prof. T. Nojima, Dr. S. Nakamura and other members of this laboratory for their kindly assistance.
\bibliography{apssamp}

\end{document}